\documentclass[12pt]{article}

\usepackage[centertags]{amsmath}
\usepackage{amsmath}\DeclareMathOperator{\sign}{sign}
\usepackage{amsfonts}
\usepackage{amssymb}
\usepackage{amsthm}
\usepackage{newlfont}
\usepackage{epsfig}
\usepackage{amscd}

\newcommand{\RR}{{\mathbb R}}

\newcommand{\beq}{\begin{equation}}
\newcommand{\eeq}{\end{equation}}
\newcommand{\ba}{\begin{array}}
\newcommand{\ea}{\end{array}}
\newcommand{\bea}{\begin{eqnarray}}
\newcommand{\eea}{\end{eqnarray}}
\newcommand{\eps}{{\epsilon}}

\begin{document}

\begin{center}
{\large \sc \bf On the dispersionless Kadomtsev-Petviashvili equation in n+1 dimensions: 
exact solutions, the Cauchy problem for small initial data 
and wave breaking}

\vskip 20pt

{\large  S. V. Manakov$^{1,\S}$ and P. M. Santini$^{2,\S}$}

\vskip 20pt

{\it 
$^1$ Landau Institute for Theoretical Physics, Moscow, Russia

\smallskip

$^2$ Dipartimento di Fisica, Universit\`a di Roma "La Sapienza", and \\
Istituto Nazionale di Fisica Nucleare, Sezione di Roma 1 \\
Piazz.le Aldo Moro 2, I-00185 Roma, Italy}

\bigskip

$^{\S}$e-mail:  {\tt manakov@itp.ac.ru, paolo.santini@roma1.infn.it}

\bigskip

{\today}

\end{center}

\begin{abstract}
We study the $n+1$-dimensional generalization of the dispersionless Kadomtsev-Petviashvili (dKP) equation, a universal  
equation describing the propagation of weakly nonlinear, quasi one dimensional waves in 
$n+1$ dimensions, and arising in several physical 
contexts, like acoustics, plasma physics and hydrodynamics. For $n=2$, this equation is   
integrable, and it has been recently shown to be a prototype model equation in the description of the 
two dimensional wave breaking of localized initial data. 
 We construct an exact solution of the $n+1$ dimensional model containing an arbitrary function of one variable, 
corresponding to its  
parabolic invariance, describing waves, constant on their paraboloidal wave front, breaking simultaneously  
in all points of it. Then we use such solution to build a uniform 
approximation of the solution of the Cauchy problem, for small and localized initial data, showing that 
such a small and localized initial data evolving according to the $n+1$-dimensional dKP equation break, 
in the long time regime, if and only 
if $1\le n\le 3$; i.e., in physical space. Such a wave breaking takes place, generically, in a point of the paraboloidal 
wave front, and 
the analytic aspects of it are given explicitly in terms of the small initial data.
 
\end{abstract}

\section{Introduction}

The $n+1$ dimensional generalization of the dispersionless Kadomtsev - Petviashvili equation:
\beq\label{KZn}
\ba{l}
\left(u_t+u u_x \right)_x+\Delta_{\bot}u=0,~~ u=u(x,\vec y,t),~~\vec y=(y_1,\dots,y_{n-1}) \\
\Delta_{\bot}=\sum\limits_{i=1}^{n-1}\partial^2_{y_i},~~n\ge 2,
\ea
\eeq
hereafter referred to as the $dKP_n$ equation, 
describes the propagation of weakly nonlinear quasi one dimensional waves in 
$n+1$ dimensions. This equation was first obtained by Timman \cite{Timman}, for $n=2$,   
in the study of unsteady motion in transonic flows, and then derived by Khoklov - Zobolotskaya, in the 3+1 
dimensional acoustic context in \cite{ZK}, where the main properties of the model were pointed out. 
Equation (\ref{KZn}) is the $x$-dispersionless limit of another distinguished model, the $n+1$ dimensional generalization 
of the Kadomtsev-Petviashvili equation \cite{KP}, integrable for $n=2$ by the Inverse Spectral Transform (IST) \cite{ZMNP,AC}. 

The universal character of (\ref{KZn}) can be explained as follows. 
Take any system of nonlinear PDEs i) characterized, for example, by nonlinearities of hydrodynamic type and 
ii) whose linear limit, 
at least in some approximation, is described by the wave equation. Then, iii) looking at the propagation of quasi - one 
dimensional waves and iv) neglecting dispersion and dissipation, one obtains, at the second order in the proper 
multiscale expansion, the $dKP_n$ equation (\ref{KZn}). Therefore (\ref{KZn}) arises in several physical 
contexts, like acoustics, plasma physics and hydrodynamics. 

Indeed, consider a nonlinear system of PDEs whose linear limit is the wave equation $f_{t't'}=\Delta f$ in $n+1$ dimensions; then  
a localized initial condition evolving according to it is concentrated, asymptotically, on the spherical wave front:
\beq
|\vec x |-t'=O(1),~~~~~~|\vec x |=\sqrt{\sum\limits_{k=1}^nx^2_k} .
\eeq
Looking for nonlinear corrections and quasi - one dimensional propagations (say, in the $x_1$ direction),  
we introduce the convenient variables:
\beq\label{KZ_variables}
x=x_1-t'=O(1),~~y_j=\eps^q x_{j+1},~j=1,\dots, n-1,~~t=\eps^{2q}t',~~q>0,
\eeq 
where $\eps$ is the order of magnitude of the small amplitude of the wave. Then   
the spherical wave front becomes, approximately, its second order contact, an ellipsoidal paraboloid:
\beq
|\vec x |-t'\sim x+\frac{1}{2t}\sum\limits_{k=1}^{n-1}y^2_k=O(1).
\eeq
In addition, starting from the dispersion relation of the wave equation and looking for quasi one dimensional waves  
in the $x_1$ direction, we have longer wave lengths (smaller wave numbers) in the transversal directions: 
$\vec k_{\bot}=\eps^q\vec \kappa_{\bot}$, obtaining that the dispersion 
relation $\omega(\vec k)$ of the wave equation reduces to that of the linearized $dKP_n$ (\ref{KZn}) (up to the 
trivial rescaling $\vec y \to \vec y /\sqrt{2}$):
\beq\label{omega_gas}
\ba{l}
\omega(\vec k)=\sqrt{\sum\limits_{j=1}^nk^2_j}\sim k_1+\eps^{2q}\frac{\kappa^2_{\bot}}{2 k_1},~~
\kappa^2_{\bot}=\sum\limits_{j=2}^n\kappa^2_j, \\
\theta(\vec x,t')=\vec k\cdot \vec x-\omega(\vec k)t' \sim k_1 x +
\vec\kappa_{\bot}\cdot \vec y -\frac{\kappa^2_{\bot}}{2 k_1}t .
\ea
\eeq 
This is what happens in the acoustic problem in $3+1$ dimensions \cite{ZK}, where the gas density, the pressure and the 
component of the velocity in the main direction $x_1$ are proportional to $\eps u(x,\vec y,t)$, and $u$ solves the 
$dKP_3$ equation.  

Starting from the plasma physics equations  
\beq
\rho_t+\nabla\cdot(\rho\vec v)=0,~~\vec v_t+(\vec v\cdot\nabla)\vec v+\nabla\phi=\vec 0,~~\Delta\phi-e^{\phi}+\rho=0 ,
\eeq
describing a gas of hot electrons on a background of cold ions, where $\rho$ and $\vec v$ are the ion density and velocity 
and $\phi$ is the electric potential \cite{DEGM}, 
the linearized theory leads instead to a generalized fourth order wave equation in ($3+1$) dimensions
\beq
\Delta\phi_{tt}=\phi_{tt}-\Delta\phi,
\eeq
reducing to the wave equation in the long wave approximation (small wave numbers):
\beq
\omega=\sqrt{\frac{k^2}{1+k^2}}\sim k(1-\frac{k^2}{2}),~~k^2=k^2_1+k^2_2+k^2_3.
\eeq

Looking, in addition, for quasi one dimensional waves (the wave numbers in the transversal directions are smaller than 
the (small) wave number in the $x_1$ direction: $k_1=\eps^p \kappa_1,~\vec k_{\bot}=\eps^{p+q}\vec \kappa_{\bot}$), we obtain
\beq
\ba{l}
\omega=\sqrt{\frac{k^2}{1+k^2}}\sim \eps^p \kappa_1+\eps^{p+2q}\frac{\kappa^2_{\bot}}{2 \kappa_1}-\eps^{3p}\frac{\kappa^3_1}{2},\\
\theta(\vec x,t')=\vec k\cdot \vec x-\omega(\vec k)t' \sim \\
\kappa_1 x +\vec\kappa_{\bot}\cdot \vec y -\left\{
\ba{cc}
\eps^{p+2q}\frac{\kappa^2_{\bot}}{2 \kappa_1}t , & p>2q, \\
\eps^{3q}\left(\frac{\kappa^2_{\bot}}{2 \kappa_1}-\frac{\kappa^3_1}{2}\right)t , & p=2q ,
\ea
\right.
\ea
\eeq
where now
\beq\label{KZ_lw_variables}
x=\eps^p(x_1-t'),~~y_j=\eps^{p+q}x_{j+1},~j=1,2,~~t=\eps^{p+2q}t' .
\eeq
If $p>2q$, the term $k^3_1$ is negligeable wrt $(\kappa^2_{\bot})/\kappa_1$ and one obtains again the dispersion relation of 
the linearized $dKP_3$ equation; if $p=2q$, they are comparable and one obtains the dispersion relation of 
the linearized KP equation in 3+1 dimensions. 

In the water wave theory, the situation is very similar to that of plasma physics and the $dKP_2$ equation is derived 
in the long wave approximation. In all the above three physical contexts, 
the choice of the exponent $q$ in (\ref{KZ_variables}),(\ref{KZ_lw_variables}) 
comes from the balance, at the second order in the proper 
multiscale expansion, with the nonlinearity of the physical system, and is $q=1/2$.
    
We remark that the $1+1$ dimensional version of (\ref{KZn}) is the celebrated Riemann-Hopf equation $u_t+u u_x=0$,  
the prototype model in the description of the gradient catastrophe (or wave breaking) of one dimensional waves \cite{W}. 
Therefore a natural question arises: do solutions of $dKP_n$ break and, if so, is it possible to give 
an analytic description of such a multidimensional wave breaking? 
 
A first and positive answer in this direction was recently given in \cite{MS0}, for the integrable   
\cite{KG,Taka,Zakharov,Kri,TT,Ferapontov} $dKP_2$ case. Indeed, using a novel IST for vector fields, we have been able 
to solve the  
Cauchy problem of $dKP_2$ \cite{MS1} and of other distinguished integrable Partial Differential Equations (PDEs) arising as  
commutation of 
multidimensional vector fields \cite{MS2,MS3,MS4}. The associated nonlinear Riemann-Hilbert (RH) inverse problem turns out to be 
an efficient tool to study several properties of the solution space of $dKP_2$, like, for instance,   
i) the construction of the longtime behaviour of the solutions of the Cauchy problem;      
ii) the possibility of establishing that localized initial profiles evolving according to dKP$_2$ break at finite time and, 
if small, 
they break in the longtime regime, investigating in an  
explicit way the analytic aspects of such a longtime wave breaking of two-dimensional waves \cite{MS0}. 
  
In this paper, motivated by the analytic results of the integrable case $n=2$ \cite{MS0}, 
we answer the above question in arbitrary dimensions, when the model (\ref{KZn}) is not integrable, 
under the assumption of small and localized initial data. 
To obtain this result, we first construct an exact solution of equation (\ref{KZn}) containing an arbitrary function 
of one variable, consequence of the   
parabolic invariance of equation (\ref{KZn}), describing a wave, constant on its paraboloidal wave front, breaking simultaneously in all points 
of it. Then we use such solution to build a uniform 
approximation of the solution of the Cauchy problem, showing that    
``small and localized initial data evolving according to the $dKP_n$ equation break, in the long time regime, if and only if  
$1\le n\le 3$; i.e., in physical space''. Such a wave breaking takes place, generically, in a point of the paraboloidal wave front, and 
the analytic aspects of it are given explicitly in terms of the initial data. In addition we show that, if the initial data are 
$O(\eps)$, then the breaking times are respectively $O(\eps^{-1})$, $O(\eps^{-2})$ and $O(e^{\eps^{-1}})$, for $n=1,2$ and $3$. 
We remark that, from the knowledge of such a breaking longtime regime of the solution, one can reconstruct exactly the initial data, an 
important issue in many physical contexts.

The existence of a critical dimensionality above which small and localized data do not break has a clear origin, since, 
in the model, two terms 
act in opposite way: the nonlinearity is responsible for the steepening of the profile, while the $n-1$ diffraction channels, 
represented by the transversal Laplacian, have an opposite effect; for $n=1,2,3$ the nonlinearity prevails 
and wave breaking takes place (but at longer and longer time scales, as $n$ increases), while, for $n\ge 4$, the number 
of transversal diffraction channels is enough to prevent such phenomenon, for small data, in the longtime regime. 

The paper is organized as follows. In \S 2 we derive the exact solutions of the model and in \S 3 we use 
them to build a uniform approximation of the solution of the Cauchy problem of (\ref{KZn}), for small and localized 
initial data, establishing that wave breaking takes place only if $n=1,2,3$. At last, in \S 4, we discuss in great detail  
the analytic aspects of such a wave breaking.

\section{Exact solutions of the $dKP_n$ equation}

The universal properties of the $dKP_n$ equation, discussed in the Introduction,  
suggest its invariance under motions on the associated paraboloid. Indeed, it is easy to show that equation 
(\ref{KZn}) admits the following Lie point symmetry group of transformations
\beq\label{Lie}
\ba{l}
\tilde x=x+\sum\limits_{i=1}^{n-1}\left(\delta_i y_i-\delta^2_i t\right),~~ \\
\tilde y_j=y_j-2\delta_j t ,~~j=1,\dots, n-1,
\ea
\eeq 
where the $\delta_j$'s are the arbitrary parameters of the group, leaving invariant the paraboloid
\beq\label{paraboloid2}
x+\frac{1}{4t}\sum\limits_{i=1}^{n-1}y^2_i=\xi .
\eeq
Correspondingly, equation (\ref{KZn}) possesses the following exact solution  
\beq\label{exactKZ}
u=\left\{
\ba{ll}
t^{-\frac{n-1}{2}}F\left(x+\frac{1}{4t}\sum\limits_{i=1}^{n-1}y^2_i-\frac{2ut}{3-n}\right), &  n\ne 3, \\
t^{-1}F\left(x+\frac{1}{4t}\sum\limits_{i=1}^{n-1}y^2_i-u~t\ln t \right), & n=3 ,
\ea
\right.
\eeq
where $F$ is an arbitrary function of one argument; such solution is characterized by the differential constraint 
$\sigma=0$, where $\sigma$ is the corresponding ``characteristic symmetry'' 
\beq
\sigma=\left(\sum\limits_{i=1}^{n-1}\delta_i y_i\right)u_x-2t\sum\limits_{i=1}^{n-1}\delta_i u_{y_i}
\eeq  
of equation (\ref{KZn}). Indeed, if one looks for solutions of (\ref{KZn}) in the form
\beq\label{v_def}
u=v(\xi,t),~~\xi=x+\frac{1}{4t}\sum\limits_{i=1}^{n-1}y^2_i,
\eeq
one obtains the following equation for $v(\xi,t)$:
\beq\label{v_equ}
v_t+\frac{n-1}{2t}v+v v_{\xi}=0.
\eeq
Its $v/t$ term can be eliminated by the change of variables  
\beq\label{q_def}
v(\xi,t)=t^{-\frac{n-1}{2}}q(\xi,\tau),
\eeq
where 
\beq\label{tau}
\ba{l}
\tau(t)=\left\{
\ba{ll}
\frac{2}{3-n}t^{\frac{3-n}{2}}, & n\ne 3, \\
\ln t, &    n=3,
\ea
\right.
\ea
\eeq 
leading to the Riemann - Hopf equation: 
\beq\label{Riemann}
q_{\tau}+q q_{\xi}=0,
\eeq
whose general solution is implicitely given by
\beq\label{sol_Riemann}
q=F(\xi-q \tau),
\eeq
where $F$ is an arbitrary function of one argument. Going back to the original variables via (\ref{v_def}), (\ref{q_def}), 
(\ref{tau}), the solution (\ref{sol_Riemann}) becomes (\ref{exactKZ}). 

We remark that, if $F$ is a regular and localized function of its argument, the solution (\ref{exactKZ}) describes a 
wave concentrated on the wave front, given by the paraboloid (\ref{paraboloid2}), and constant on it, breaking,   
simultaneously, on the whole paraboloid. We also remark that, for $n=2$, the solution (\ref{exactKZ}) has been first derived 
using a recently developed method to construct exact solutions of the nonlinear RH inverse problem associated with the 
integrable $dKP_2$ equation \cite{MS5}.  
 
\section{The Cauchy problem and wave breaking of small and localized initial data}

Since the paraboloid (\ref{paraboloid2}) plays an important role in the asymptotics of the $dKP_n$ equation (see the 
Introduction), the exact solution (\ref{exactKZ}) is physically relevant and can be used to build a uniform 
approximation of the solution of the Cauchy problem for the $dKP_n$ equation, under the 
hypothesis of small and localized initial data. 

The basic idea (already used in \cite{Manakov2}, in a different context) is that, if the initial condition is small:
\beq\label{initial_u}
u(x,\vec y,0)=\eps u_0(x,\vec y),~~0<\eps < < 1,
\eeq
the solution of the Cauchy problem for $dKP_n$ is well approximated by the corresponding solution for the 
linearized $dKP_n$ untill one enters the nonlinear regime, in which the Riemann - Hopf equation (\ref{Riemann}) 
becomes relevant, eventually causing wave breaking. Since the breaking time of $O(\eps)$ initial data evolving 
according to (\ref{Riemann}) is $\tau(t)=O(\eps^{-1})$, the nonlinear regime for $dKP_n$ is characterized by the condition 
$t=O(\tau^{-1}(\eps^{-1}))$, where $\tau^{-1}$ is the inverse of (\ref{tau}); so that:
\beq
\tau^{-1}\left(\eps^{-1}\right)=\left\{
\ba{cc}
\eps^{-\frac{2}{3-n}} & \mbox{if } 1\le n < 3, \\
e^{\eps^{-1}} & \mbox{if }n=3 ,
\ea
\right.
\eeq
and a proper matching has to be made between the solution of the linearized $dKP_n$, valid for $t< < O(\tau^{-1}(\eps^{-1}))$, 
and the exact solution of the previous section, valid in the nonlinear regime $t=O(\tau^{-1}(\eps^{-1}))$. 
\subsection{The linear regime}
Since the initial condition (\ref{initial_u}) is small, the solution of $dKP_n$ is well approximated, for finite times, 
by the solution of the linearized $dKP_n$ equation:
\beq\label{sol_linear}
u(x,\vec y,t)\sim \frac{\eps}{(2\pi)^n} 
\int_{\RR^n}\hat u_0(k_1,\vec k_{\bot})e^{i(k_1 x+\vec k_{\bot}\cdot\vec y-\frac{k^2_{\bot}}{k_1}t)}dk_1~d\vec k_{\bot}
\eeq
where $\hat u_0(k_1,\vec k_{\bot})$ is the Fourier transform of $u_0(x,\vec y)$
\beq
\hat u_0(k_1,\vec k_{\bot})=\int_{\RR^n}u_0(x,\vec y)e^{-i(k_1 x+\vec k_{\bot}\cdot\vec y)}dx d\vec y.
\eeq
Such approximation is also valid in the longtime interval:
\beq\label{region2}
1< < t < < O\left(\tau^{-1}\left( \eps^{-1}\right)\right) 
\eeq
(far away from the nonlinear regime), in which the solution of $dKP_n$ is described by the standard stationary phase 
approximation of the multiple integral (\ref{sol_linear}):
\beq\label{stationary_phase}
\ba{l}
u(x,\vec y,t)\sim t^{-\frac{n-1}{2}}\eps G\left(x+\frac{1}{4t}\sum\limits_{i=1}^{n-1}y^2_i,\frac{\vec y}{2t}\right), 
\ea
\eeq
where
\beq\label{def_G}
\ba{l}
G\left(\xi ,\vec \eta\right):=2^{-n}\pi^{-{\frac{n+1}{2}}}\int_{\RR}d\lambda |\lambda |^{{\frac{n-1}{2}}}
\hat u_0(\lambda,\lambda\vec\eta)e^{i\lambda \xi -i\frac{\pi}{4}(n-1) \sign\lambda} , 
\ea
\eeq
valid in the space-time region (\ref{region2}) and 
\beq
(x-\xi )/t,~y_i/t=O(1), ~~~i=1,\dots,n ,
\eeq
on the paraboloid (\ref{paraboloid2}). Outside the paraboloid, the solution decays faster.   
This formula says that the localized initial condition (\ref{initial_u}) has evolved concentrating, asymptotically, 
on the paraboloid (\ref{paraboloid2}).
\subsection{The nonlinear regime}
The approximate solution of $dKP_n$ in the nonlinear regime $t=O(\tau^{-1}(\eps^{-1}))$, obtained matching equations 
(\ref{stationary_phase}) and (\ref{exactKZ}), reads as follows:
\beq\label{nonlinear}
u(x,\vec y,t)\sim u^{as}_n(x,\vec y,t)\equiv \left\{
\ba{ll}
t^{-\frac{n-1}{2}}\eps G\left(x+\frac{1}{4t}\sum\limits_{i=1}^{n-1}y^2_i-\frac{2ut}{3-n},\frac{\vec y}{2t}\right), & n\ne 3, \\
t^{-1}\eps G\left(x+\frac{1}{4t}\sum\limits_{i=1}^{n-1}y^2_i-u~t\ln t,\frac{\vec y}{2t} \right), &  n=3. 
\ea
\right.
\eeq

Since the term ($ut$) inside the first argument of function $G$, responsible for the wave breaking, is $O(t^{\frac{3-n}{2}})$ 
for $n\ne 3$, and the analogous term $ut\ln t$ is 
$O(\ln t)$ for $n=3$, then these terms are large for $n=1,2,3$ and infinitesimal for $n\ge 4$. 
It follows that wave breaking takes place only for $n=1,2,3$; 
for $n\ge 4$ the solution (\ref{nonlinear}) coincides with the linearized solution (\ref{stationary_phase}), 
and no breaking takes place. 
We remark that, in the integrable case $n=2$, one recovers the results obtained in \cite{MS0} using the IST for 
vector fields. We also remark that, if the breaking regime (\ref{nonlinear}) is known (measured), i.e., if 
function $G$ is known, the initial condition $\eps u_0(x,\vec y)$ is uniquely reconstructed simply inverting (\ref{def_G}).

It is possible to show that the error made approximating the solution of $dKP_n$ by  
(\ref{nonlinear}) is given by $u=u^{as}_n(x,\vec y,t)(1+O(t^{-1}))$ for $n=2,3$.                          

Summarizing, the asymptotic solution (\ref{nonlinear}) illustrates the following breaking picture for the $dKP_n$ 
equation (\ref{KZn}), 
corresponding to localized and $O(\eps)$ initial data. If $n=1$ (the Riemann - Hopf case), waves break in the 
longtime regime $t=O(\eps^{-1})$; 
if $n=2$, waves break in the longtime regime $t=O(\eps^{-2})$, much later than in the $1+1$ dimensional case;  
also if $n=3$ small waves break, but at an exponentially large time scale: $t=O(e^{\frac{1}{\eps}})$; at last, if $n\ge 4$, 
small and localized initial data do not break in the longtime regime. This result 
has a clear physical meaning: increasing the dimensionality of the transversal space, the number of diffraction 
channels of the wave increases, untill such diffraction, acting for a long time, is strong enough to prevent the gradient 
catastrophe 
of the small $n$ dimensional wave. It is a remarkable coincidence that small initial data break, in the longtime regime, 
only in $1+1$, $2+1$ and $3+1$ dimensions; i.e., only in physical space! 

We end this section remarking that, if the initial condition is the Gaussian $u_0(x,\vec y)=d_n~exp(-\frac{x^2+|\vec y|^2}{4})$, 
where $d_n$ is a constant, then the above asymptotic solution can be written in terms of elementary or special functions, 
depending on $n$:  
\beq\label{Gn}
\ba{l}
G(\xi,\vec\eta)=\frac{d_n}{\sqrt{\pi}}\frac{1}{(1+|\vec\eta |^2)^{\frac{n+1}{4}}}
\left[\cos\frac{\pi (n-1)}{4}
\Gamma\left(\frac{n+1}{4}\right)~_1 F_1\left(\frac{n+1}{4},\frac{1}{2};-\frac{Y^2}{4}\right)+\right. \\
\left. \sin\frac{\pi (n-1)}{4}
\Gamma\left(\frac{n+3}{4}\right) Y~_1 F_1\left(\frac{n+3}{4},\frac{3}{2};-\frac{Y^2}{4}\right)\right], \\
Y:=\frac{\xi}{\sqrt{1+|\vec\eta |^2}} ,
\ea
\eeq
where $\Gamma$ is the Euler Gamma function and$~_1 F_1$ is the Kummer confluent hypergeometric function \cite{AS}. 
If, in particular, $n=2$ and $n=3$, with $d_2=\sqrt{2 \pi}$ and $d_3=2$, one obtains, respectively,
\beq\label{G2}
\ba{l}
G(\xi,\eta)=(1+\eta^2)^{-\frac{3}{4}}
\left[\Gamma\left(\frac{3}{4}\right)~_1 F_1\left(\frac{3}{4},\frac{1}{2};-\frac{Y^2}{4}\right)+ \right. \\
\left. Y~ \Gamma\left(\frac{5}{4}\right)~_1 F_1\left(\frac{5}{4},\frac{3}{2};-\frac{Y^2}{4}\right)\right]
\ea
\eeq
(see Figure 1) and
\beq\label{G3}
\ba{l}
G(\xi,\eta_1,\eta_2)=\frac{\xi}{(1+\eta^2_1+\eta^2_2)^{\frac{3}{2}}}e^{-\frac{\xi^2}{4(1+\eta^2_1+\eta^2_2)}} .
\ea
\eeq
\begin{center}
\mbox{ \epsfxsize=12cm \epsffile{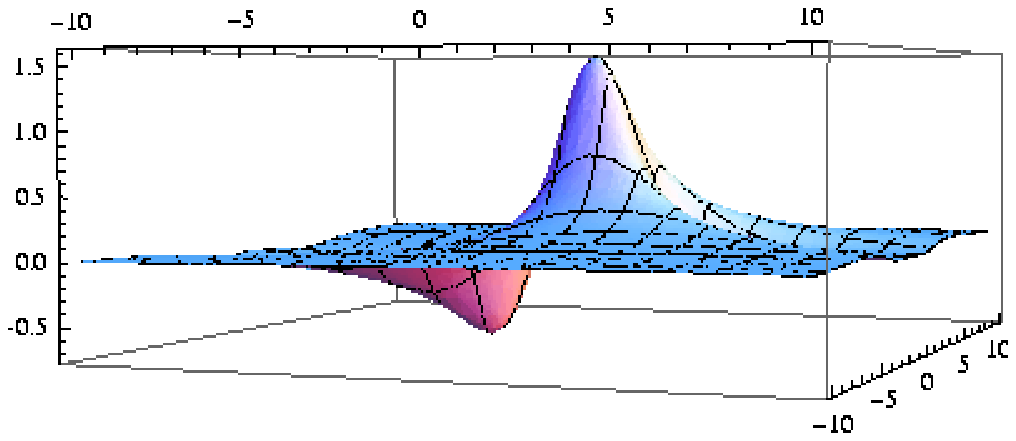}}
\end{center}
Figure 1. The plotting of function (\ref{G2}).  

\section{Geometric and analytic aspects of the wave breaking}

In this section we show how to derive, from the asymptotics (\ref{nonlinear}), the analytic features of the wave breaking for 
$n=2,3$, in terms of the initial data, represented by function $G$ defined in (\ref{def_G}). 
We first rewrite equations (\ref{nonlinear}) in the more convenient form
\beq\label{sol_Hopf}
\ba{l}
w\sim \eps G(\zeta,\vec \eta), \\
\xi=\eps G(\zeta,\vec \eta)\tau +\zeta,
\ea
\eeq
where
\beq\label{transf_n=2}
w=\sqrt{t}u,~~\tau=2\sqrt{t},~~\xi=x+\frac{1}{4t}y^2,~~\eta=\frac{y}{2t}
\eeq
for $n=2$, and 
\beq\label{transf_n=3}
w=t u,~~\tau=\ln t,~~\xi=x+\frac{1}{4t}(y^2_1+y^2_2),~~\vec\eta=\frac{\vec y}{2t}
\eeq
for $n=3$, describing the evolution of an $n$-dimensional wave according to the $1+1$ dimensional Riemann - Hopf equation 
$w_{\tau}+ww_{\xi}=0$. In the following, we mainly concentrate on the case $n=3$; the case $n=2$, that can be easily 
recovered setting to zero all the partial derivatives of $G$ with respect to $\eta_2$ in the formulae of this section, 
has been discussed in detail in \cite{MS0}.   

One solves the second of equations (\ref{sol_Hopf}) 
with respect to the parameter $\zeta$, obtaining $\zeta(\xi,\vec\eta,\tau),$ 
and replaces it into the first, to obtain the solution  
$w\sim \eps G(\zeta(\xi,\vec\eta,\tau),\vec\eta)$. The inversion of the second of equations (\ref{sol_Hopf}) 
is possible iff its $\zeta$-derivative is different from zero. Therefore the singularity manifold (SM) of 
the solution  is the $n$ - dimensional manifold characterized by the equation 
\beq\label{SM_Hopf}
{\cal S}(\zeta,\vec\eta,\tau)\equiv 1+\eps G_{\zeta}(\zeta,\vec\eta)\tau=0~~~
\Rightarrow~~~\tau=-\frac{1}{\eps G_{\zeta}(\zeta,\vec\eta)}.
\eeq
Since 
\beq
\label{gradw}
\ba{l}
\nabla_{(\xi,\vec\eta)}w=\frac{\eps \nabla_{(\zeta,\vec\eta)}G(\zeta,\vec\eta)}{1+\eps G_{\zeta}(\zeta,\vec\eta)\tau},
\ea
\eeq
the slope of the localized wave becomes infinity (the so-called gradient catastrophe) on the SM, 
and the $n$-dimensional wave ``breaks''. 
 
Then the first breaking time $\tau_b$ and the corresponding characteristic parameters  
${\vec{\zeta}}_b=({\zeta}_b,\vec\eta_b)$ are defined by the global minimum of a function of $n$ variables: 
\beq
\label{tilde_t_b}
\tau_b=-\frac{1}{\eps G_{\zeta}({\vec {\zeta}}_b)}=
\mbox{global min}\left(-\frac{1}{\eps G_{\zeta}(\zeta,\vec\eta)}\right)>0,
\eeq 
and it is characterized, together with the condition $G_{\zeta}(\vec {\zeta}_b)<0$, 
by the condition that the symmetric quadratic form 
$<\underline z,\cal{H} \underline z>$ be positive $\forall\underline z\in\RR^n $, where $\cal{H}$ 
is the Hessian matrix of function $G_{\zeta}(\zeta,\vec\eta)$, evaluated at 
$\vec \zeta=\vec {\zeta}_b=(\zeta_b,\vec\eta_b)$. 

The corresponding point at which the first wave breaking 
takes place is, from (\ref{sol_Hopf}), ${\vec \xi}_b=(\xi_b,\vec\eta_b)\in\RR^n$, where:
\beq\label{tilde_x_b}
\xi_b= {\zeta}_b + \eps G({\vec {\zeta}}_b)\tau_b . 
\eeq 

Now we evaluate equations (\ref{sol_Hopf}) and (\ref{SM_Hopf}) near breaking, in the regime:
\beq
\label{near_breaking_Hopf}
\xi=\xi_b+\xi',~~\vec\eta=\vec\eta_b+\vec\eta',~~\tau=\tau_b+\tau',~~\zeta={\zeta}_b+{\zeta'},
\eeq
where $\xi',\vec\eta',\tau',\zeta'$ are small. Using (\ref{tilde_t_b}) - (\ref{tilde_x_b}), the second of  
equations (\ref{sol_Hopf}) becomes, at the leading order, 
the following cubic equation in ${\zeta'}$:
\beq\label{cubic}
{\zeta'}^3+a(\vec\eta'){\zeta'}^2+b(\vec\eta',\tilde\tau){\zeta'}-\gamma X(\xi',\vec\eta',\tilde\tau)=0,
\eeq
where
\beq\label{def_a,b,X}
\ba{l}
a(\vec\eta')=\frac{3}{G_{\zeta\zeta\zeta}}(G_{\zeta\zeta\eta_1}\eta'_1+G_{\zeta\zeta\eta_2}\eta'_2), \\
b(\vec\eta',\tilde\tau)=
\frac{3}{G_{\zeta\zeta\zeta}}\left[2 G_{\zeta}\tilde\tau +
G_{\zeta\eta_1\eta_1}{\eta'_1}^2+2G_{\zeta\eta_1\eta_2}\eta'_1\eta'_2+G_{\zeta\eta_2\eta_2}{\eta'_2}^2 \right], \\
X(\xi',\vec\eta',\tilde\tau)=\xi'-\eps G({\zeta}_b,\vec\eta_b+\vec\eta')\tau'-
\eps \left[G({\zeta}_b,\vec\eta_b+\vec\eta')-G\right]\tau_b  \sim \\
\xi'+(\frac{G_{\eta_1}}{G_{\zeta}}\eta'_1+\frac{G_{\eta_2}}{G_{\zeta}}\eta'_2)-
\frac{G}{|G_{\zeta}|} \tilde\tau+
\frac{1}{2G_{\zeta}}(G_{\eta_1\eta_1}{\eta'_1}^2+2G_{\eta_1\eta_2}\eta'_1\eta'_2+ \\
G_{\eta_2\eta_2}{\eta'_2}^2) -\frac{1}{|G_{\zeta}|}(G_{\eta_1}\eta'_1+G_{\eta_2}\eta'_2)\tilde\tau +
\frac{1}{6G_{\zeta}}(G_{\eta_1\eta_1\eta_1}{\eta'_1}^3+ \\
3 G_{\eta_1\eta_1\eta_2}{\eta'_1}^2\eta'_2+3 G_{\eta_1\eta_2\eta_2}\eta'_1{\eta'_2}^2+G_{\eta_2\eta_2\eta_2}{\eta'_2}^3),~~~~
\gamma=\frac{6|G_{\zeta}|}{G_{\zeta\zeta\zeta}},
\ea
\eeq
and  
\beq\label{epsilon}
\tilde\tau \equiv \frac{\tau'}{\tau_b}=\frac{\tau -\tau_b}{\tau_b},
\eeq
corresponding to the maximal balance
\beq\label{max_balance_Hopf}
|{\zeta'}|,|\eta'_1|,|\eta'_2|=O(|\tilde\tau |^{1/2}),~~~|X|=O(|\tilde\tau |^{3/2}).
\eeq
In (\ref{def_a,b,X}) and in the rest of this section, all partial derivatives of $G$ whose 
arguments are not indicated are meant to be evaluated at $\vec {\zeta}_b=(\zeta_b,\vec\eta_b)$.

The three roots of the cubic are given by the well-known Cardano-Tartaglia formula:
\beq\label{sol_cubic1}
\ba{l}
{\zeta'}_1\left(\xi',\vec\eta',\tilde\tau\right)=-\frac{a}{3}+(A_+)^\frac{1}{3} + (A_-)^\frac{1}{3} , \\
{\zeta'}_{\pm}\left(\xi',\vec\eta',\tilde\tau\right)=-\frac{a}{3}-\frac{1}{2}\left((A_+)^\frac{1}{3} + 
(A_-)^\frac{1}{3}\right) \pm \frac{\sqrt{3}}{2}i\left((A_+)^\frac{1}{3} - (A_-)^\frac{1}{3}\right),  
\ea
\eeq
where
\beq\label{def_A}
\ba{l}
A_{\pm}=R\pm \sqrt{\Delta}
\ea
\eeq
and the discriminant $\Delta$ reads
\beq
\label{discriminant}
\Delta=R^2+Q^3,
\eeq
with
\beq
\label{def_QR_Hopf} 
\ba{l}
Q(\vec\eta',\tilde\tau)=\frac{3b-a^2}{9}=-\frac{2 |G_{\zeta}|}{G_{\zeta\zeta\zeta}}\tilde\tau + 
\frac{1}{G^2_{\zeta\zeta\zeta}}\Big[(G_{\zeta\zeta\zeta}G_{\zeta\eta_1\eta_1}-G^2_{\zeta\zeta\eta_1}){\eta'_1}^2+ \\
+2(G_{\zeta\zeta\zeta}G_{\zeta\eta_1\eta_2}-G_{\zeta\zeta\eta_1}G_{\zeta\zeta\eta_2})\eta'_1\eta'_2+ 
(G_{\zeta\zeta\zeta}G_{\zeta\eta_2\eta_2}- \\
G^2_{\zeta\zeta\eta_2}){\eta'_2}^2\Big],       \\
R(\xi',\vec\eta',\tilde\tau)=\frac{\gamma}{2}X(\xi',\vec\eta',\tilde\tau)+\frac{ab}{18}+
\frac{a}{3}Q(\vec\eta',\tilde\tau).
\ea
\eeq
At the same order, function $\cal S$ in (\ref{SM_Hopf}) becomes
\beq
\label{SM_Hopf_2}
\ba{l}
{\cal S}(\zeta',\vec\eta',\tilde\tau)=-\tilde\tau+
\frac{1}{2|G_{\zeta}|}\Big[G_{\zeta\zeta\zeta}{\zeta'}^2+G_{\zeta\eta_1\eta_1}{\eta'_1}^2+
G_{\zeta\eta_2\eta_2}{\eta'_2}^2+ \\
2G_{\zeta\zeta\eta_1}{\zeta'}\eta'_1+2G_{\zeta\zeta\eta_2}{\zeta'}\eta'_2+
2G_{\zeta\eta_1\eta_2}\eta'_1\eta'_2\Big]= \\
-\tilde\tau + \frac{1}{2|G_{\zeta}|}<\vec{\xi'},\cal{H}\vec{\xi'}>,~~~\vec{\xi'}=(\xi',\vec\eta').
\ea
\eeq
Known ${\zeta'}$ as function of ($\xi',\vec\eta',\tilde\tau$) from the cubic (\ref{cubic}), 
the solution $w$ and its gradient are then approximated, near breaking, 
by the formulae:
\beq
\label{V_x}
\ba{l}
w(\xi,\vec\eta,\tau)\sim \eps G({\zeta}_b+{\zeta'},\vec\eta_b+\vec\eta'),    \\
\nabla_{(\xi,\vec\eta)}w\sim \eps \frac{\nabla_{(\zeta',\vec\eta')}G({\zeta}_b+{\zeta'},\vec\eta_b+\vec\eta')}
{{\cal S}(\zeta',\vec\eta',\tilde\tau)}.
\ea
\eeq

\subsubsection{Before breaking} 

If $\tau<\tau_b$ $(\tilde\tau <0)$, the coefficient $Q$ in (\ref{def_QR_Hopf}) is strictly positive, 
due to the positivity of the Hessian quadratic form; 
then the discriminant $\Delta=R^2+Q^3$ is also strictly positive  
and only the root ${\zeta'}_1$ is real. Correspondingly, the real solution $w$ is 
single valued and described by Cardano's formula (see Figure 2). In addition, function $\cal S$ in (\ref{SM_Hopf_2}) 
is also strictly positive and $\nabla_{(\xi,\vec\eta)}w$ is finite $\forall ~\xi,\vec\eta$. 

To have a more explicit solution, we first restrict the asymptotic region 
to a narrower volume, so that the cubic (\ref{cubic}) reduces to the linear equation $b\zeta'=\gamma X$; then the solution 
exhibits a universal behaviour, coinciding with the following 
exact similarity solution of equation $w_{\tau}+ww_{\xi}=0$:  
\beq
\label{simil_asymp_1}
\ba{l}
w\sim 
\frac{\xi-\xi_b +(G_{\eta_1}/G_{\zeta})(\eta_1-{\eta_1}_b)+
(G_{\eta_2}/G_{\zeta})(\eta_2-{\eta_2}_b)}{\tau-\tau_b}=\frac{\vec\nu\cdot (\vec\xi-\vec\xi_b)}{\tau-\tau_b}, 
\ea
\eeq
describing the hyperplane tangent to the wave 
, where 
\beq
\label{k}
\ba{l}
{\vec \nu}=\left(1,\frac{G_{\eta_1}}{G_{\zeta}},\frac{G_{\eta_2}}{G_{\zeta}}\right)
\ea
\eeq
defines the breaking direction. In addition, in such a narrow volume: 
\beq
\nabla_{(\xi,\vec\eta)}w\sim \frac{1}{\tau-\tau_b}\vec\nu .
\eeq
We also look at the different balance: $\xi',\eta'_j$ of the same order, and $\tilde \tau \le O(|\eta'_j|)$, 
suitable for taking the $\tau \uparrow \tau_b$ limit. In this case the cubic simplifies to ${\zeta'}^3\sim\gamma X$ and 
\beq\label{befbreak1}
\ba{l}
w\sim \eps G\left({\zeta}_b+\sqrt[3]{\gamma X(\xi',\vec\eta',\tilde\tau)},\vec\eta_b+ \vec\eta'\right)~\Rightarrow~ \\
\nabla_{\xi,\vec\eta}w\sim \frac{1}{3}\sqrt[3]{\frac{6|G_{\zeta}|}{G_{\zeta\zeta\zeta}}}
\frac{\eps \nabla_{\zeta,\vec\eta}G}{\sqrt[3]{\left( X(\xi',\vec\eta',\tilde\tau)\right)^2}}.
\ea
\eeq

\subsubsection{At breaking} 

As $\tau \uparrow \tau_b$,  
the above tangent hyperplane (now tangent at the breaking point) has an infinite slope and equation 
$G_{\zeta}\xi'+G_{\eta_1}\eta'_1+G_{\eta_2}\eta'_2=0$.

At the breaking time $\tau=\tau_b$ one can give an explicit description of the 
vertical inflection. If $|\xi'/\eta'_1|,~|\xi'/\eta'_2|=O(1)$, the cubic (\ref{cubic}) simplifies 
to ${\zeta'}^3=\gamma X$ and equation (\ref{befbreak1}) becomes
\beq\label{break1}
\ba{l}
w\sim \eps G\left({\zeta}_b+
\sqrt[3]{\gamma X_b},\vec\eta_b+\vec\eta' \right)~\Rightarrow~ \nabla_{\xi,\vec\eta}w\sim \frac{\sqrt[3]{\gamma}}{3}
\frac{\eps \nabla_{\zeta,\vec\eta}G}{\sqrt[3]{{X_b}^2}}
\ea
\eeq
where $X_b(\xi',\vec\eta')\equiv X(\xi',\vec\eta',0)$ is defined in (\ref{def_a,b,X}). We remark that 
$X_b\sim \xi'+(\frac{G_{\eta_1}}{G_{\zeta}}\eta'_1+\frac{G_{\eta_2}}{G_{\zeta}}\eta'_2)$, 
if $\xi'=\alpha_1\eta'_1+\alpha_2\eta'_2,~\alpha_j\ne -G_{\eta_j}/G_{\zeta},~j=1,2$, while  
$X_b\sim 
\frac{1}{2G_{\zeta}}(G_{\eta_1\eta_1}{\eta'_1}^2+2G_{\eta_1\eta_2}\eta'_1\eta'_2+ G_{\eta_2\eta_2}{\eta'_2}^2) $
if $\alpha_j= -G_{\eta_j}/G_{\zeta},~j=1,2$.  

Equation (\ref{break1}) implies that, if $n=2$, all the derivatives of $w$ blow up at $\tau=\tau_b$, in 
the breaking point $(\xi_b,\eta_b)$, with the universal law $X^{-2/3}_b$, except the derivative along the transversal line 
$X_b(\xi',\eta')=0$, 
represented by the vector field $\hat X={X_b}_{\eta}\partial_{\xi}-\partial_{\eta}$, for which
\beq
\hat X w |_{(\xi_b,\eta_b)}=-\eps G_{\eta}.
\eeq
If $n=3$, the situation is similar: all derivatives of $w$ blow up at $\tau=\tau_b$, in 
the breaking point $(\xi_b,\vec\eta_b)$, with the universal law $X^{-2/3}_b$, except the derivatives along the transversal 
surface $X_b(\xi',\vec\eta')=0$, 
having as natural basis the vector fields $\hat X_j={X_b}_{\eta_j}\partial_{\xi}-\partial_{\eta_j},~j=1,2$, for which:
\beq
\hat X_j w |_{(\xi_b,\vec\eta_b)}=-\eps G_{\eta_j},~j=1,2.
\eeq  

\subsubsection{After breaking}

After breaking, the solution becomes three-valued in a compact region of the $(\xi,\vec\eta)$ - space (see Figures 2, 3, 4), 
and does not describe any physics;   
nevertheless a detailed study of the multivalued region is important, in view of a proper regularization 
of the model. 

If $\tau > \tau_b$ $(\tilde\tau >0)$, in the regime (\ref{max_balance_Hopf}), the SM equation ${\cal S}=0$:  
\beq\label{ellipse_Hopf}
\ba{l}
2|G_{\zeta}|\tilde\tau=
G_{\zeta\zeta\zeta}{\zeta'}^2+G_{\zeta\eta_1\eta_1}{\eta'_1}^2+
G_{\zeta\eta_2\eta_2}{\eta'_2}^2+ \\
2G_{\zeta\zeta\eta_1}{\zeta'}\eta'_1+2G_{\zeta\zeta\eta_2}{\zeta'}\eta'_2+
2G_{\zeta\eta_1\eta_2}\eta'_1\eta'_2
\ea
\eeq
describes an ellipsoidal paraboloid in the ($\zeta',\vec\eta',\tilde t$) space, with minimum at 
the breaking point $(\vec\xi_b,\tilde\tau_b)$.  

Eliminating $\zeta'$ from equations (\ref{ellipse_Hopf}) and (\ref{cubic}), one obtains the SM equation in space-time 
coordinates,
coinciding with the $\Delta =0$ condition, where $\Delta$, $Q$ and $R$ are defined in (\ref{discriminant}) and 
(\ref{def_QR_Hopf}). 

For $n=2$, the SM is a closed curve with two cusps in the $(\xi,\eta)$ - plane at the points
\beq\label{cusps} 
\pm \sqrt{\frac{2 |G_{\zeta}|G_{\zeta\zeta\zeta}}{G_{\zeta\zeta\zeta}G_{\zeta\eta\eta}-G^2_{\zeta\zeta\eta}}}
(\frac{G_{\eta}}{|G_{\zeta}|},1)\sqrt{\tilde\tau}
\eeq
(see Figure 3), corresponding to the conditions $Q=R=0$, on which the three real 
solutions of the cubic coincide. On the remaining part of the closed curve $\Delta=0$, two of the three real branches coincide: 
$w_1\sim \eps G({\zeta}_b+\zeta'_1,\vec \eta),~w_{+}=w_{-}\sim \eps G(\zeta_b+\zeta'_{+},\vec \eta)$. Outside the closed curve, 
$\Delta>0$ and the real solution $w$ is single valued; inside the closed surface 
$\Delta <0$ and the real solution $w$ is three valued. We remark that the transversal and longitudinal widths of such a closed 
curve are respectively $O(\tilde\tau^{\frac{1}{2}})$ 
and $O(\tilde\tau^{\frac{3}{2}})$. Therefore this curve develops, at $\tau=\tau_b$, 
from the breaking point $(\xi_b,\eta_b)$, with an infinite speed in the tranversal direction, and with zero 
speed in the longitudinal direction, recovering the results obtained in \cite{MS0}.  

For $n=3$, the SM is a closed surface in the $(\xi,\vec\eta)$ - space 
made of two surfaces having the same boundary: the transversal closed curve $Q=R=0$ (see Figure 4), on which the three real 
solutions of the cubic coincide.  The $Q=0$ condition defines an ellipse in the $(\eta_1,\eta_2)$ plane with 
semi-axes $\sqrt{\frac{2|G_{\zeta}|G_{\zeta\zeta\zeta}}{\lambda_{\pm}}}\sqrt{\tilde\tau}$, where
\beq
\ba{l}
\lambda_{\pm}=\frac{a_1+a_2\pm\sqrt{(a_1-a_2)^2+4 c^2}}{2}, \\
a_j=G_{\zeta\zeta\zeta}G_{\zeta\eta_j\eta_j}-G^2_{\zeta\zeta\eta_j},~j=1,2, ~~~
c=G_{\zeta\zeta\zeta}G_{\zeta\eta_1\eta_2}-G_{\zeta\zeta\eta_1}G_{\zeta\zeta\eta_2}.
\ea
\eeq  
As before, in the remaining part of the closed curve $\Delta=0$, two of the three real branches 
coincide. While the axes of the transversal closed curve $Q=R=0$ are of $O(\tilde\tau^{\frac{1}{2}})$, the thickness of the longitudinal 
region between the two surfaces is of $O(\tilde\tau^{\frac{3}{2}})$. Therefore this closed surface develops, at $\tau=\tau_b$, 
from the breaking point $(\xi_b,\vec\eta_b)$, with an infinite speed in the tranversal directions, and with zero 
speed in the longitudinal direction. Intersecting this closed surface with any plane containing the $\xi$ - axis, 
one obtains a closed curve with two cusps as in Figure 3; therefore the closed curve $Q=R=0$ is made of all these cusps.   

We end these considerations remarking that the similarity solution before breaking, 
the vertical inflection at breaking, and the compact three-valued region after breaking 
make clear the universal character of the gradient catastrophe of two- and three-dimensional waves evolving 
according to the Riemann-Hopf equation (see Figure 2).

\vskip 10pt
Since the transformations (\ref{transf_n=2}),(\ref{transf_n=3}) are globally invertible, for $t\ne 0$:
\beq
n=2:~~~u=\frac{1}{\sqrt{t}}w(\xi,\eta,\tau),~~t=\frac{\tau^2}{4},~~x=\xi-\frac{\eta^2\tau^2}{4},~~y=\frac{\eta\tau^2}{2},
\eeq
\beq
n=3:~~~u=\frac{1}{t}w(\xi,\vec\eta,\tau),~~t=e^{\tau},~~~~x=\xi-e^{\tau}(\eta^2_1+\eta^2_2),~~\vec y=2e^{\tau}\vec\eta,
\eeq 
all the above considerations are easily transfered to the $dKP_n$ case. In particular, small and localized initial data 
evolving according to the $dKP_n$ equation (\ref{KZn}) break, at $t_b=\tau^2_b/4$ in the point 
$(x_b,y_b)=(\xi_b-\tau^2_b\eta^2_b/4,\eta_b\tau^2_b/2)$ of the parabolic wave front $x+y^2/(4 t_b)=\xi_b$ if $n=2$ \cite{MS0}, and 
at $t_b=e^{\tau_b}$ in the 
point $(x_b,\vec y_b)=(\xi_b-e^{\tau_b}|\vec\eta_b|^2,2e^{\tau_b}\vec\eta_b)$ of the paraboloidal wave front 
$x+({y^2_1}+{y^2_2})/(4 t_b)=\xi_b$ if $n=3$. In addition, all the previous considerations concerning the universal 
character of such a wave breaking: the similarity solution before breaking, 
the vertical inflection at breaking, and the compact three-valued space region after breaking, are transfered in a 
straightforward way to the $dKP_n$ equation, for $n=2,3$.

For instance, at $t=t_b$ and in the space region $|x'+{\eta_1}_by'_1+{\eta_2}_by'_2|=O(|y'_j|/(2 t_b))$ (the transformed of 
$|\xi'|=O(|\eta'_j|)$) equations (\ref{break1}) become 
\beq\label{break1a}
\ba{l}
u\sim \eps t^{-\frac{n-1}{2}}_b G\left({\zeta}_b+\sqrt[3]{\gamma \tilde X_b (x',\vec y')},\vec\eta_b+\frac{1}{2 t_b}\vec y' \right)
~~~~\Rightarrow~ \\
\nabla_{(x,\vec y)}u \sim \eps t^{-\frac{n-1}{2}}_b \frac{\sqrt[3]{\gamma}}{3 \sqrt[3]{\tilde X^2_b}}
\left(G_{\zeta},G_{\zeta}\vec\eta_b +\frac{1}{2 t_b}\nabla_{\vec\eta}G \right),
\ea
\eeq
where $x'=x-x_b,~\vec y'=\vec y-\vec y_b$ and $\tilde X_b(x',\vec y')=X_b(\xi',\vec\eta')$. 
Again, if $n=2$, all derivatives of $u$ at the breaking point $(x_b,y_b)$ blow up, except that along the transversal 
line $\tilde X_b(x',y')=0$, for which: 
\beq
({\tilde X}_y u_x -u_y)|_{(x_b,y_b)}=-\frac{\eps}{2 {t_b}^{3/2}}G_{\eta}. 
\eeq
Analogously, if $n=3$, all derivatives of $u$ at the breaking point $(x_b,\vec y_b)$ blow up, except those along the 
transversal surface $\tilde X_b(x',\vec y')=0$, represented by the basis vector fields 
$\hat {\tilde X}_j=\tilde {X_b}_{y_j}\partial_x-\partial_{y_j},~j=1,2$, for which:
\beq
\hat {\tilde X}_j u|_{(x_b,\vec y_b)}=-\frac{\eps}{2 {t_b}^{2}}G_{\eta_j},~j=1,2.
\eeq

\newpage

\begin{center}
\mbox{ \epsfxsize=6cm \epsffile{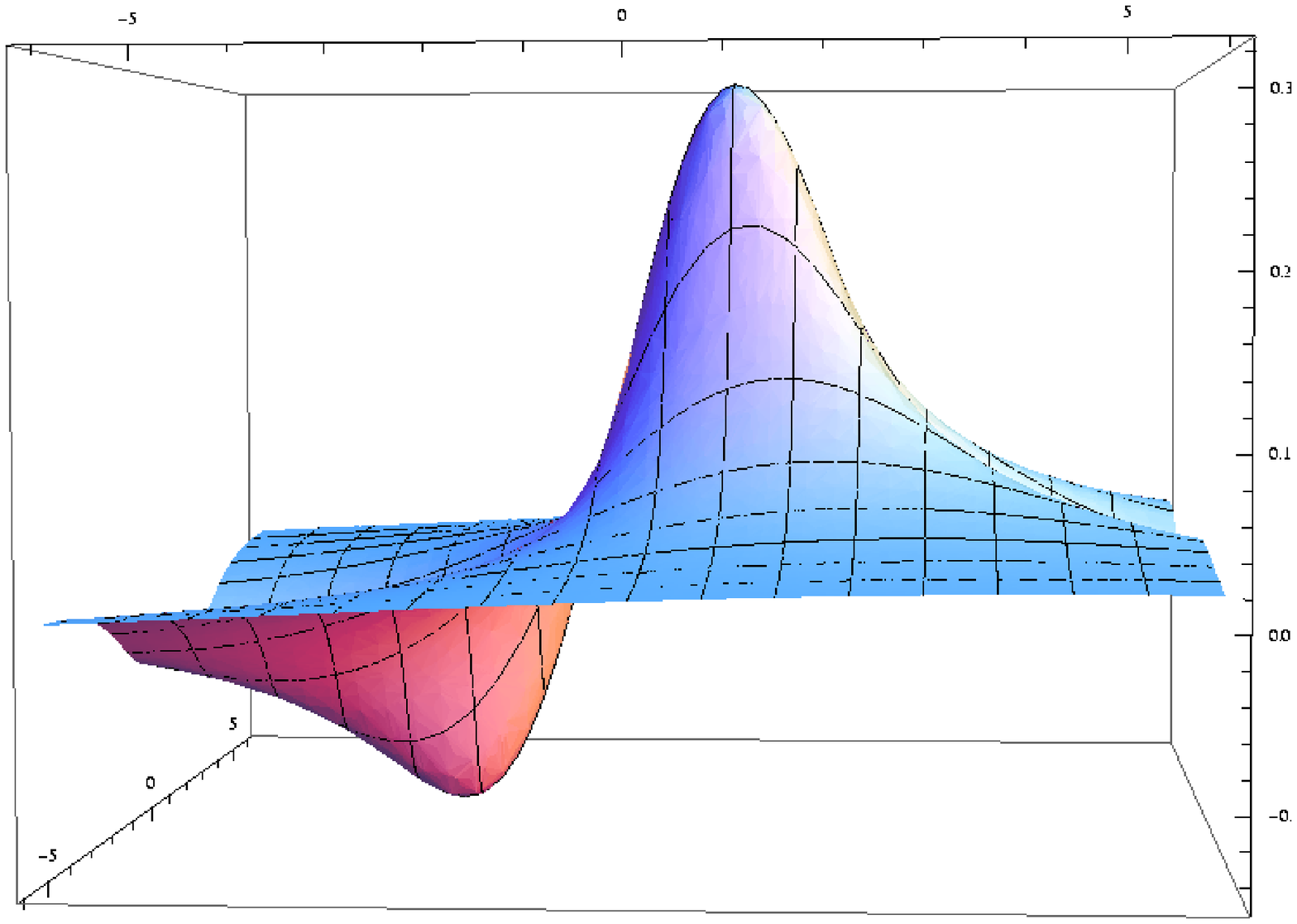}}
\mbox{ \epsfxsize=6cm \epsffile{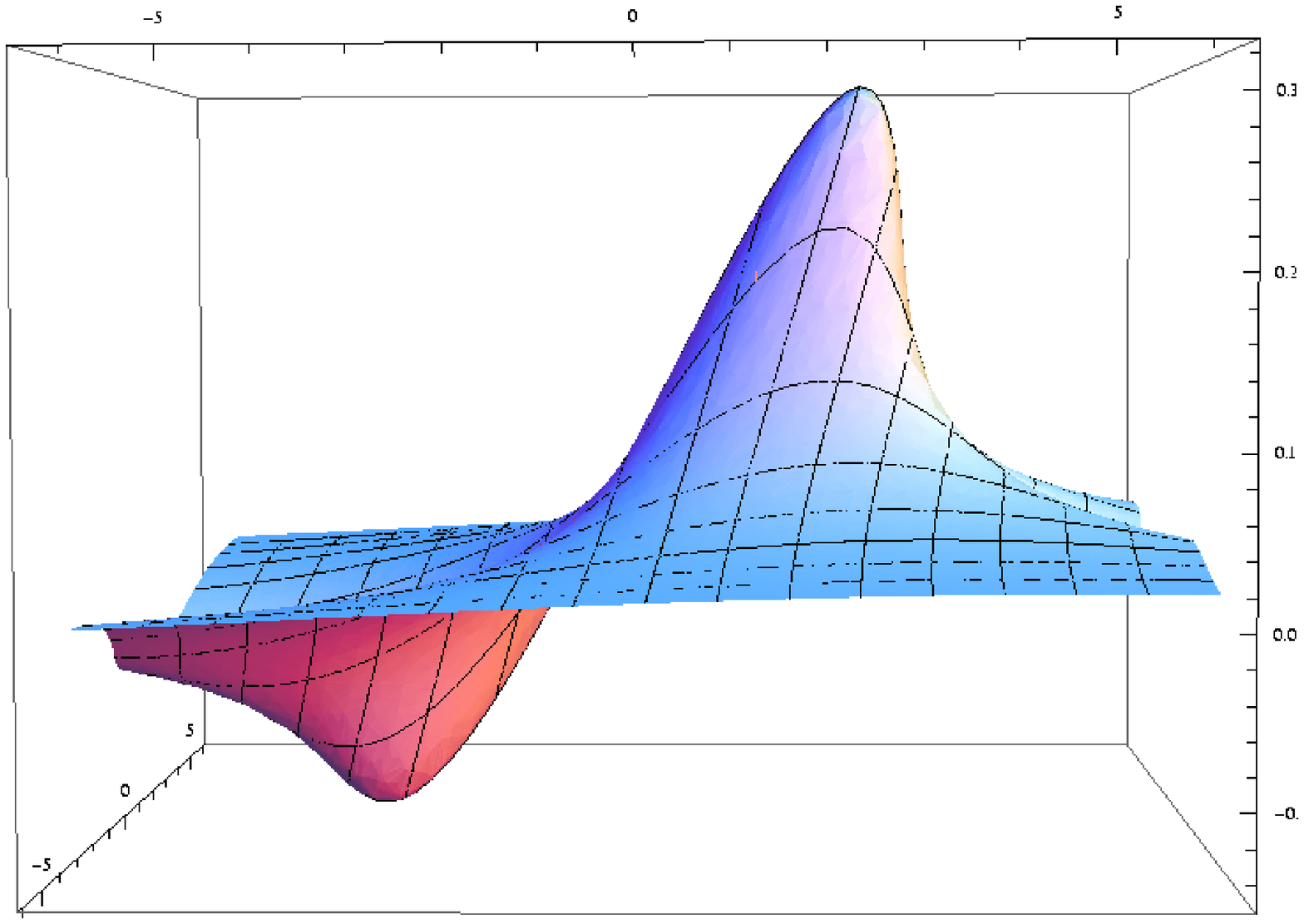}}
\mbox{ \epsfxsize=6cm \epsffile{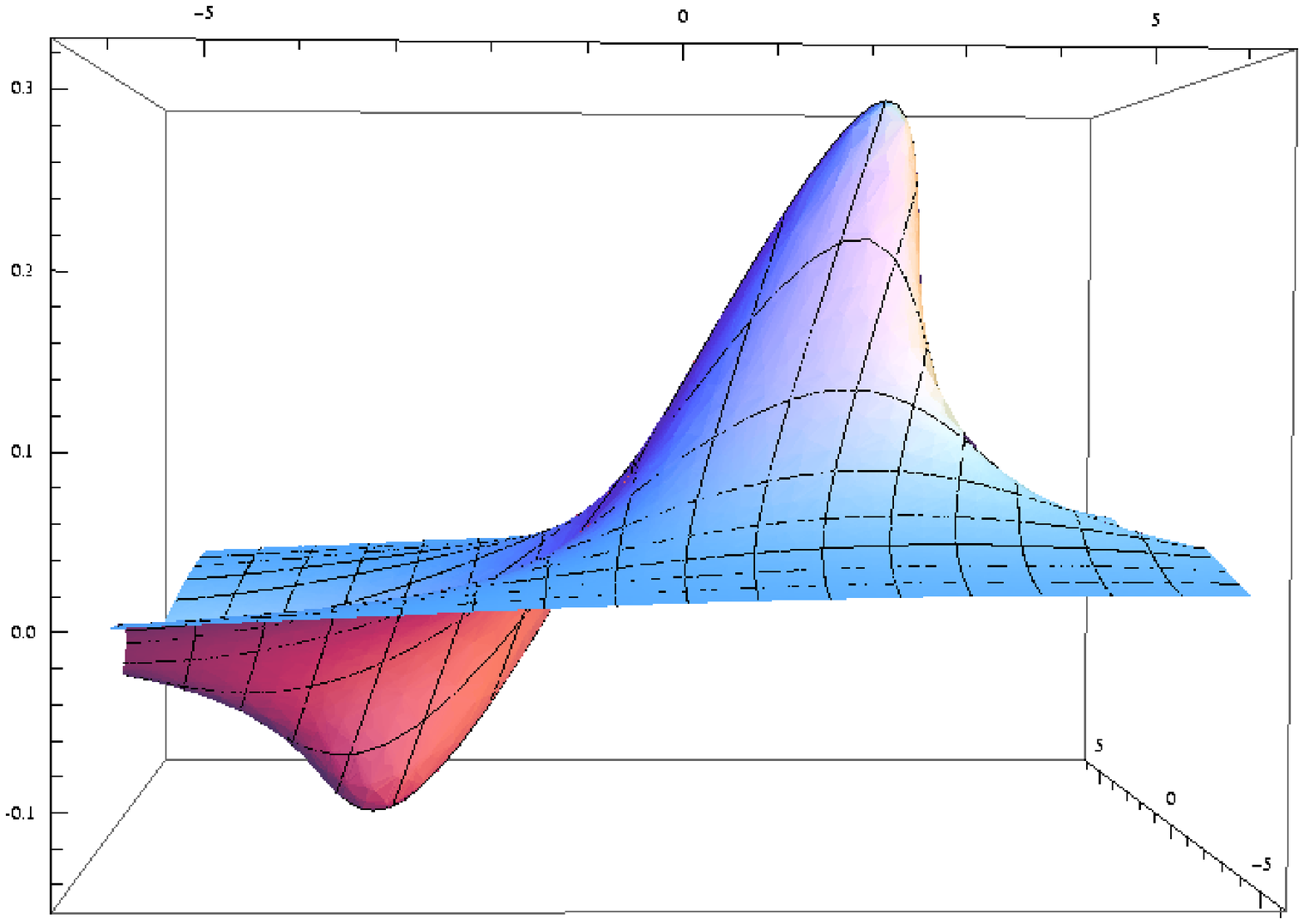}}
\mbox{ \epsfxsize=6cm \epsffile{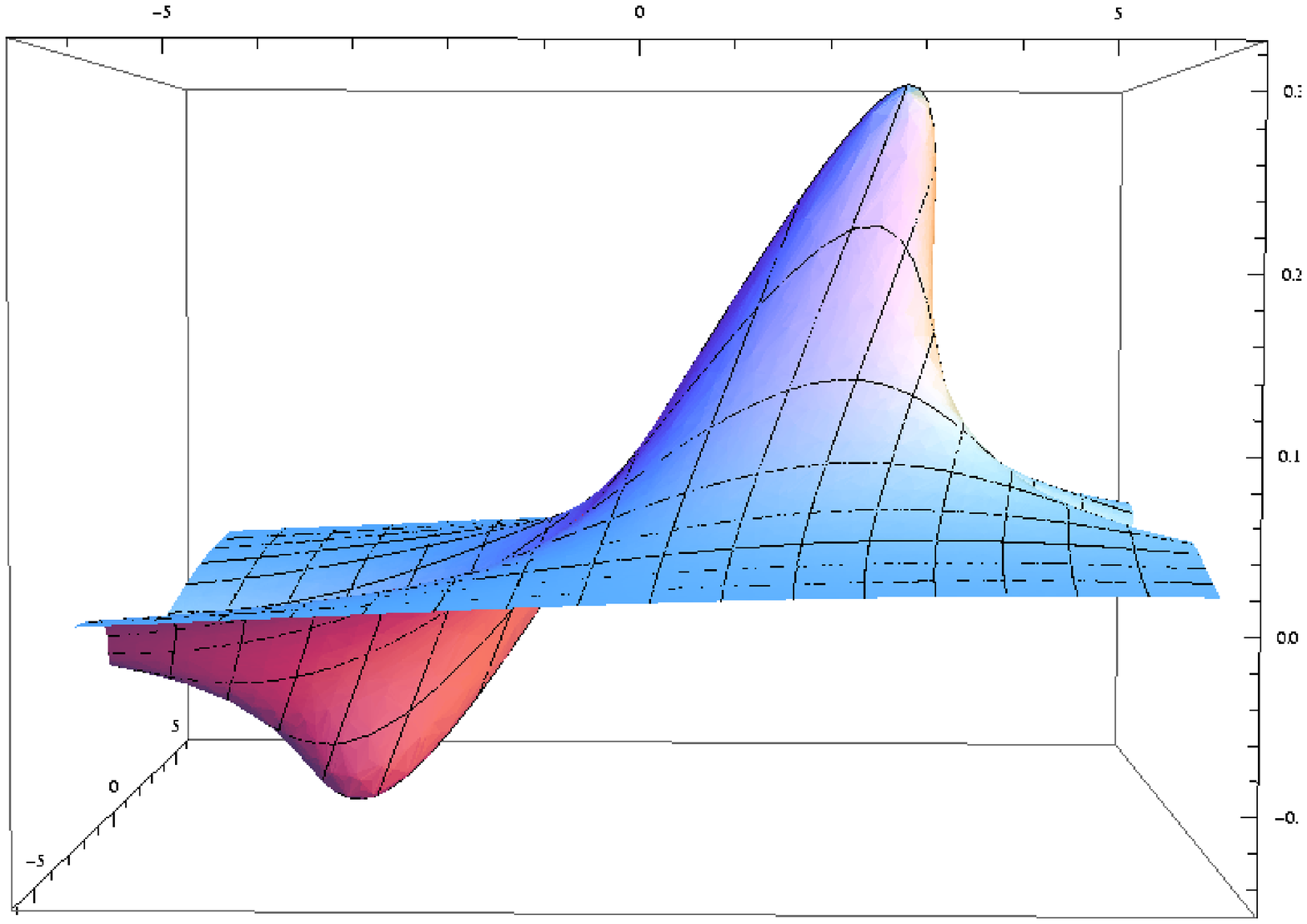}}
\end{center} 
Figure 2. For $n=2$, $\eps=0.2$, and $G(\xi,\eta)$ given by (\ref{G2}), four consecutive snapshots, at $\tau=0$, 
$\tau=\tau_b-1$, $\tau=\tau_b$ and $\tau=\tau_b+1$, where $\tau_b\sim 6.57$, for the evolution described by (\ref{sol_Hopf}). 
\newpage
\begin{center}
\mbox{ \epsfxsize=6cm \epsffile{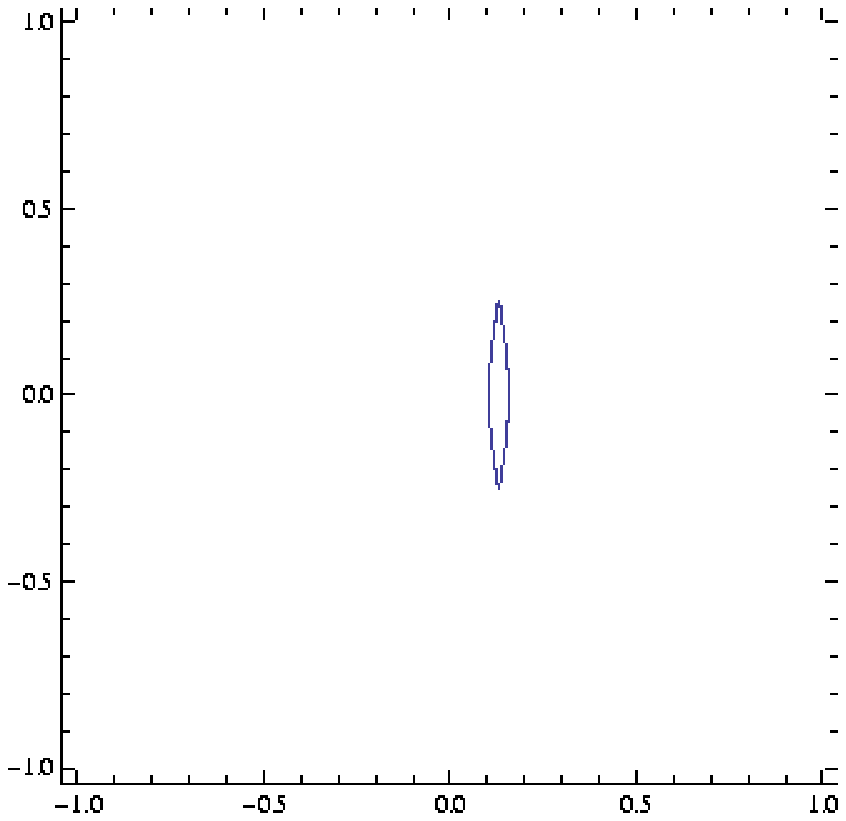}}$~~~~$
\mbox{ \epsfxsize=6cm \epsffile{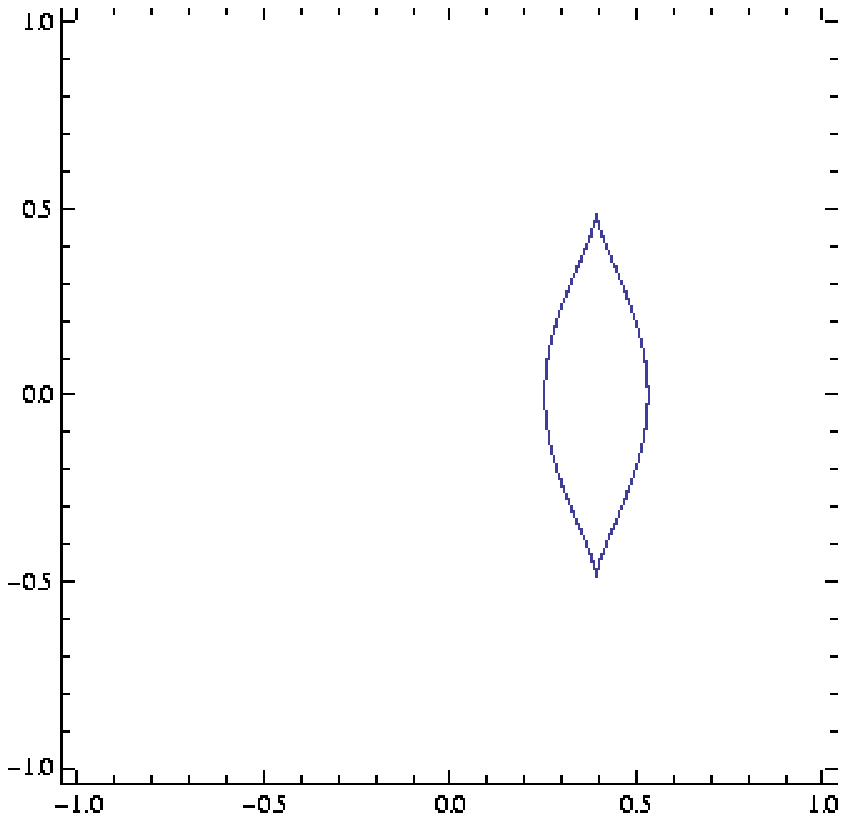}}
\end{center}
Fig. 3. For $n=2$ and $G(\zeta,\eta)$ given by (\ref{G2}), two consecutive snapshots, at $\tilde\tau=0.1$ and 
$\tilde\tau=0.3$, describing the evolution of the three-valued region, from the breaking point $(\xi_b,\eta_b)\sim (3.26,0)$ 
(the center of the figures). This region is delimited by a closed curve with two cusps.

\begin{center}
\mbox{ \epsfxsize=6cm \epsffile{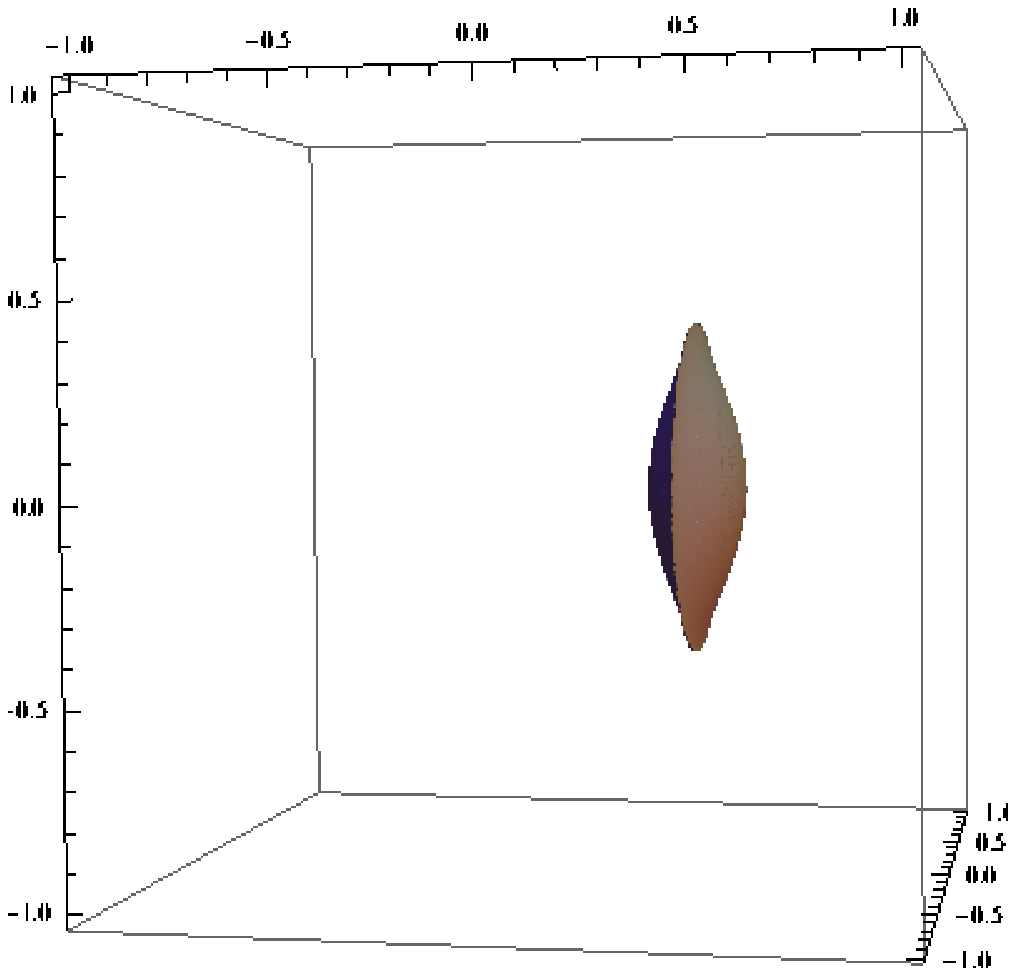}}$~~~~$
\mbox{ \epsfxsize=6cm \epsffile{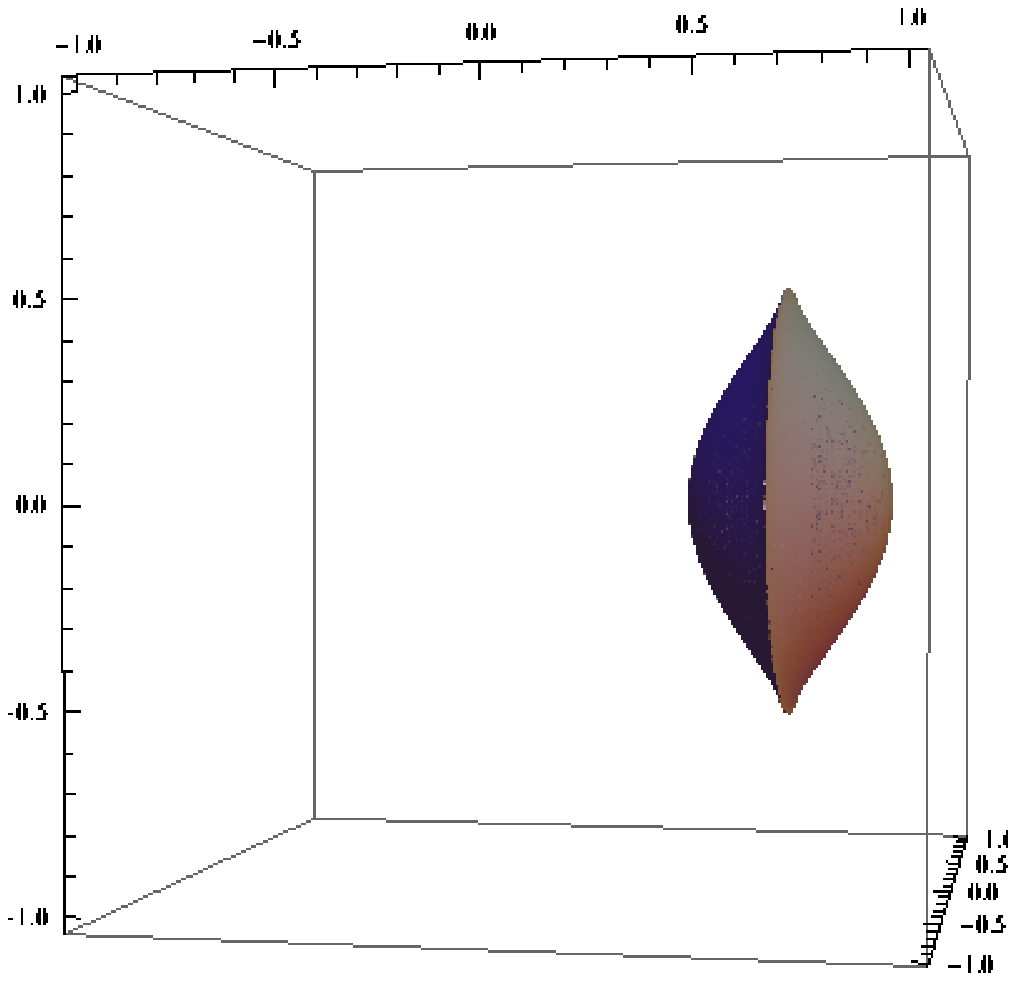}}
\end{center} 

\noindent
Figure 4. For $n=3$ and $G(\zeta,\eta_1,\eta_2)$ given by (\ref{G3}), two consecutive snapshots, at $\tilde\tau=0.3$ and 
$\tilde\tau=0.5$, describing the evolution of the compact three-valued region from the breaking 
point (the center of the figures).   
The closed surface delimiting the three-valued region is made of two surfaces with the same boundary, 
the closed curve $Q=R=0$. 

\begin{center}
\mbox{ \epsfxsize=10cm \epsffile{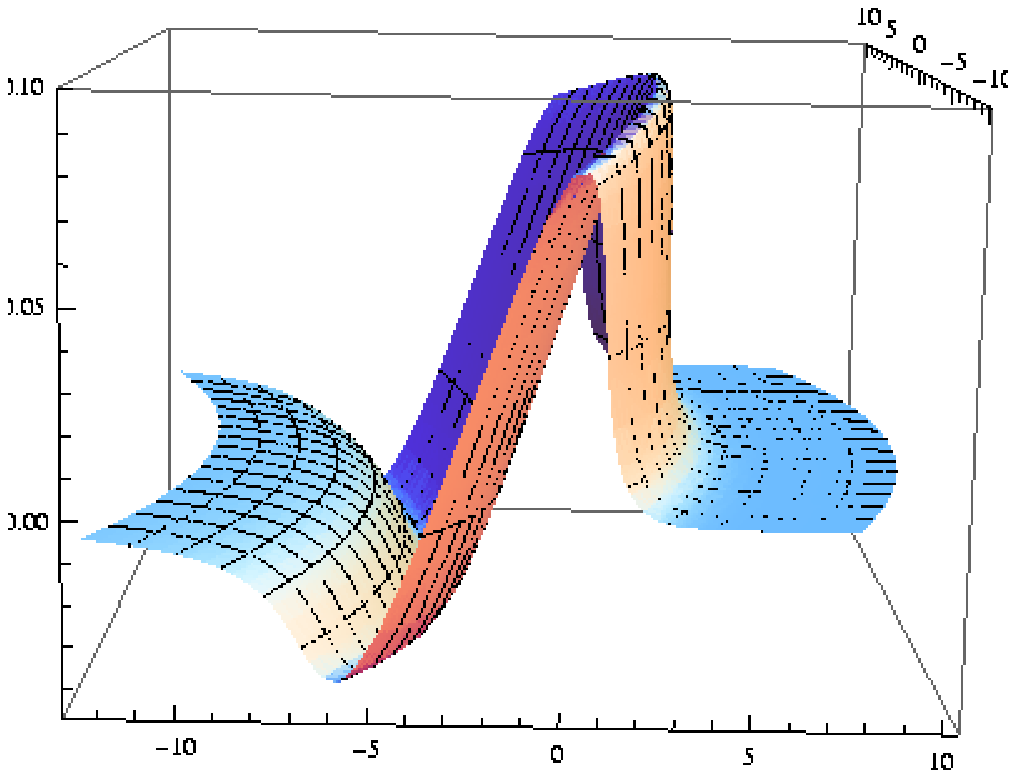}}
\end{center}
Fig. 5. For $n=2$, $\eps=0.2$ and $G(\zeta,\eta)$ given by (\ref{G2}), a detail of the parabolic wave front 
of $dKP_2$ at $t=t_b$, around the breaking point ($x_b,y_b$).
 
\vskip 5pt
\noindent
{\bf Acknowledgements}. This research has been supported by the RFBR 
grants 07-01-00446, 08-01-90104, and 09-01-92439, by the bilateral agreement between the Consortium Einstein 
and the RFBR, and by the bilateral agreement between the 
University of Roma ``La Sapienza'' and the Landau Institute for Theoretical Physics of the 
Russian Academy of Sciences.  


\begin{thebibliography}{9}

\bibitem{Timman} R. Timman, ``Unsteady motion in transonic flow'', Symposium Transsonicum, Aachen 1962. 
Ed. K. Oswatitsch,  Springer 394-401.

\bibitem{ZK} E. A. Zobolotskaya and R. V. Kokhlov, ``Quasi - plane waves in the nonlinear 
acoustics of confined beams'',  Sov. Phys. Acoust. {\bf 15}, n. 1 (1969) 35-40. 

\bibitem{KP} B. B. Kadomtsev and V. I. Petviashvili, ``On the stability of solitary waves in weakly 
dispersive media'', Sov. Phys. Dokl., {\bf 15}, 539-541 (1970).

\bibitem{ZMNP} V. E. Zakharov, S. V. Manakov, S. P. Novikov and L. P. Pitaevsky, {\it Theory of solitons}, 
Plenum Press, New York, 1984. 

\bibitem{AC} M. J. Ablowitz and P. A. Clarkson, 
{\it Solitons, nonlinear evolution equations and Inverse Scattering}, London Math. Society Lecture Note 
Series, vol. 194, Cambridge University Press, Cambridge (1991).

\bibitem{DEGM} R. K. Dodd, J. C. Eilbeck, J. D. Gibbon, H. C. Morris, {\it Solitons and nonlinear wave equations}, 
Academic Press, 1982.

\bibitem{W} J. B. Witham, {\it Linear and nonlinear waves}, J. Wiley and sons, New York, 1974.

\bibitem{MS0} S. V. Manakov and P. M. Santini: ``On the solutions of the dKP equation: the nonlinear 
Riemann-Hilbert problem, longtime behaviour, implicit solutions and wave breaking''; J. Phys. A: Math. 
Theor. {\bf 41} (2008) 055204 (23pp).

\bibitem{KG} Y. Kodama and J. Gibbons, ``Integrability of the dispersionless KP hierarchy'', 
Proc. 4th Workshop on Nonlinear and Turbulent Processes in Physics, World Scientific, Singapore 1990.

\bibitem{Taka} K. Takasaki, ``Quasi-classical limit of BKP hierarchy and $W$-infinity symmetries''.  
Lett. Math. Phys.  28  (1993),  no. 3, 177--185. 

\bibitem{Zakharov} V. E. Zakharov, ``Dispersionless limit of integrable systems in 2+1 
dimensions'', in {\it Singular Limits of Dispersive Waves}, edited by N. M. Ercolani et al., 
Plenum Press, New York, 1994.

\bibitem{Kri} I. M. Krichever, ``The $\tau$-function of the universal Witham hierarchy, matrix 
models and topological field theories'', Comm. Pure Appl. Math. {\bf 47}, 437-475 (1994). 

\bibitem{TT} K. Takasaki and T. Takebe, ``Integrable hierarchies and dispersionless limit'', 
Rev. Math. Phys. {\bf 7}, 743-808 (1995).

\bibitem{Ferapontov} E. V. Ferapontov and K. R. Khusnutdinova: ``On integrability of 
(2+1)-dimensional quasilinear systems'', Comm. Math. Phys. {\bf 248} (2004) 187-206.

\bibitem{MS1} S. V. Manakov and P. M. Santini: ``The Cauchy problem on the plane for the 
dispersionless Kadomtsev-Petviashvili equation''; JETP Letters, {\bf 83}, No 10, 462-466 (2006).
http://arXiv:nlin.SI/0604016.

\bibitem{MS2} S. V. Manakov and P. M. Santini: ``Inverse scattering problem for 
vector fields and the Cauchy problem for the heavenly equation'',   
Physics Letters A {\bf 359} (2006) 613-619.

\bibitem{MS3} S. V. Manakov and P. M. Santini: ``A hierarchy of integrable PDEs in 
$2+1$ dimensions associated with $1$ - dimensional vector fields''; Theor. Math. Phys. {\bf 152}(1), 1004-1011 (2007).
  
\bibitem{MS4} S. V. Manakov and P. M. Santini: ``The dispersionless 2D Toda equation: dressing, 
Cauchy problem, longtime behaviour, implicit solutions and wave breaking'', J. Phys. A: Math. Theor. {\bf 42} 
(2009) 095203 (16pp).

\bibitem{MS5} S. V. Manakov and P. M. Santini: ``Solvable vector nonlinear Riemann problems, exact implicit 
solutions of dispersionless PDEs and wave breaking'', arXiv:1011.2619. (J. Phys. A: Math. Theor., submitted to).

\bibitem{Manakov2} V. G. Kamensky and S. V. Manakov, ``Formation of instability regions from unstable states in 
nonlinear systems with dissipation''.  Nonlinear evolutions (Balaruc-les-Bains, 1987),  531-535, World Sci. Publ., 
Teaneck, NJ, 1988.

\bibitem{AS} Abramowitz and Stegun: {\it Handbook of Mathematical Functions with Formulas, Graphs, and Mathematical 
Tables}, Dover Publications, New York (1972).

\end{thebibliography}
\end{document}